# On the Role of Software Architecture in DevOps Transformation: An Industrial Case Study

Mojtaba Shahin [a], M. Ali Babar [b]
[a] Faculty of IT, Monash University, Australia
[b] School of Computer Science, University of Adelaide, Australia
mojtaba.shahin@monash.edu, ali.babar@adelaide.edu.au

## ABSTRACT

Development and Operations (DevOps), a particular type of Continuous Software Engineering, has become a popular Software System Engineering paradigm. Software architecture is critical in succeeding with DevOps. However, there is little evidence-based knowledge of how software systems are architected in the industry to enable and support DevOps. Since architectural decisions, along with their rationales and implications, are very important in the architecting process, we performed an industrial case study that has empirically identified and synthesized the key architectural decisions considered essential to DevOps transformation by two software development teams. Our study also reveals that apart from the chosen architecture style, DevOps works best with modular architectures. In addition, we found that the performance of the studied teams can improve in DevOps if operations specialists are added to the teams to perform the operations tasks that require advanced expertise. Finally, investment in testing is inevitable for the teams if they want to release software changes faster.

## 1 Introduction

The emergence of Development and Operations (DevOps) paradigm as a promising approach to build and release software at an accelerated pace has stimulated widespread industrial interests [1]. However, establishing DevOps culture (e.g., shared responsibility) and implementing its practices such as Continuous Delivery and Deployment (CD) require new organizational capabilities and innovative techniques and tools for some, *if not all*, Software Engineering (SE) activities [2] [3] [4].

There has been an increasing amount of literature on DevOps, which mostly deals with integrating security [5] [6], enhancing test and deployment automation [7] [8] [9], and improving performance [10] in DevOps pipelines. Other studies report the challenges (e.g., limited visibility of customer environments) that organizations encountered in DevOps adoption and the practices (e.g., test automation) employed to address those challenges [11] [12] [13] [14]. Another line of research concerns the required changes in organizational structures and developers' skills and responsibilities for adopting DevOps [15] [16] [17].

It is becoming important to understand how an application should be (re-) architected to support DevOps [1, 3]. Software Architecture (SA) is expected to be the keystone for reaching the highest level of DevOps success [2, 4]. Most of the reported research on SA and DevOps relatedness has been conducted in the context of CD as a key practice of DevOps [18] [19] [20]. In our previous paper [18], we focused on the perception of practitioners from different organizations around the world on how SA being impacted by or impacting CD, resulting in a conceptual framework to support the process of (re-) architecting for CD. Microservices are recently being leveraged as a promising architectural style to enable DevOps [21] [22] [23] [24].

Whilst DevOps, CD, and microservices share common characteristics (e.g., automating repetitive tasks) and reinforce each other [25, 26], organizations may adopt only one of the practices to achieve their business goals, e.g., delivering quality software in a shorter time more reliably [27]. Furthermore, some of the essential characteristics of CD appear to be incompatible with the constraints imposed by an organizational domain [14] [28]. What is more, the practical applicability of architectural practices or tactics for enabling and supporting DevOps is tightly associated with organizational context (i.e., domain), and the research around software architectures that enable and support DevOps (here referred to as DevOps-driven software architectures) is in its infancy [4]. This deserves a fair, thorough, specific, and contextual investigation of different architectural tactics, practices, and challenges for DevOps, as argued by Zhu et al. [29], this helps to understand "*which (architectural) practices are best for which kinds of systems in which kinds of organizations?*".

We believe that thoroughly analyzing and understanding those architectural decisions, along with their rationales and implications, that fall under DevOps is an essential step towards characterizing DevOps-driven software architectures. To achieve this goal, we carried out an exploratory case study with two teams in a company, henceforth referred to as the case company, to investigate the following research question:

**Research Question.** What are the key architectural decisions in DevOps transformation?

Our study has identified and synthesized eight architectural decisions, along with their rationales and implications, that were essential to DevOps transformation in the studied teams. Our study has also revealed that apart from the chosen architecture style, DevOps works best with modular architectures. Finally, our study identifies two concrete improvement areas for the studied teams: (i) operations specialists should be leveraged to manage shared infrastructures among teams and to perform the operations tasks that require advanced expertise, and (ii) investment in testing could be a significant driver for releasing software more quickly.

**Paper organization**: Section 2 outlines our research method. We present our findings in Section 3. Section 4 discusses the lessons learned from our work, and we then discuss the threats to our study's validation in Section 5. Section 6 examines the related work. Finally, we close the paper in Section 7.

## 2 Research Design

Given DevOps is a relatively new phenomenon and the exploratory nature of our research question, we applied a case study approach to gaining a deep understanding of the role of software architecture in DevOps transformation in the context of a company [30]. A case study is "an empirical method aims at investigating contemporary phenomena in their context" [31]. Our case study was an exploratory, holistic, two-case study, as we studied two teams from the same company [30, 32]. Informed by the established guidelines for conducting a case study [30, 31], a research protocol was developed in advance and was strictly followed when performing the case study.

### 2.1 Context

**The Case Company**. The case company is a research and development organization, which develops and delivers robust Big Data solutions and technologies (e.g., tools). By providing such Big Data capability, the customers and end users of the case company are enabled to make critical decisions faster and more accurately. The case company is made up of several teams working on various research and development programs. Each team includes a variety of roles, such as software engineers, software architects, and data scientists. In this study, we studied two teams: **TeamA** and **TeamB**.

**TeamA**. TeamA develops a social media monitoring platform that collects the available online multimedia data (text, image, and video) and tries to make them digestible to security analysts. This can enable the analysts to extract and identify intelligence and unforeseen insights quickly. Facebook and Twitter are the main data sources for this platform. This project is to descriptively summarize social media content by applying image processing, and Natural Language Processing (NLP) approaches. TeamA consists of 8 members, including software engineers, developers, and software architects in a cross-functional team. The team started by four members for about 18 months, but by growing the project, more people were added. The platform is a greenfield project. TeamA started with microservices architecture style to have multiple independent deployment units at production, but they changed the architecture of its platform to deploy one monolithic deployment unit at production.

**TeamB**. TeamB is another team in the case company that works on a greenfield platform. The platform aims at identifying and tracking the potential social event trends. The platform ingests a large amount of publicly available data from a diverse range of social media websites (e.g., Facebook). The goal is to automatically and accurately predict and track society level events such as protests, celebrations, and disease outbreaks. The data science team will then use the predictions. TeamB consists of two teams: one engineering team and one data science team. The work style is that the data science team is the customer for the engineering team, and the data science team has its own customers (e.g., security analysts). The engineering team is composed of 5 members, including system architect and software engineers. The team had recently re-architected its platform from a monolith to a new architecture style (i.e., the team calls it as *micro architecture*), to more rapidly introduce new data sources into the platform.

### 2.2 Data Collection

The data collection process initially started by an informal meeting with CTO of the case company and two key informants at the case company. The informal meeting enabled us to get a basic understanding of the case company's structure and domain. It also helped us to find the teams adopting DevOps to be used as a reference point for further steps of our case study. Furthermore, the team members who were suitable for interviews (e.g., those who had a broad view of the software development process, such as software architects and senior software engineers) were identified during the meeting. Finally, the meeting helped us to understand what documents and artifacts in the case company should be investigated.

**Table 1**. Projects, teams and interviewees

| Team | Project | Team Size | Interviewees | |
|------|---------|-----------|--------------|---|
| | | | Role (ID) | IT Exp. |
| TeamA | Social Media Platform | 8 | Software Engineer (**PA1**) | 10yr. |
| | | | Solution Architect (**PA2**) | 15yr. |
| | | | Software Architect (**PA3**) | 20yr. |
| TeamB | Social Media Platform | 9 | Software Engineer (**PB1**) | 6yr. |
| | | | System Architect (**PB2**) | 12yr. |
| | | | Software Engineer (**PB3**) | 2yr. |

Face-to-face, semi-structured interviews were the primary tool of data collection. We conducted six interviews in total, three interviews with each team. From TeamA, one software engineer, one solution architect, and one software architect, with an average of 15 years of experience in the IT industry, participated in the interviews (See Table 1). We also interviewed two (senior) software engineers and one system architect in TeamB, who had an average of 6.6 years of experience in the IT industry (See Table 1). Each interview had 30 open-ended questions, but we asked follow-up questions based on the participants' responses. The initial questions in the interviews were demographic (e.g., participants' experiences in their current role). The next questions asked about the team's organization and the characteristics of the project (e.g., domain, deployment frequency, tools, and technologies used for deployment pipeline). For example, we asked, "*Could you describe your team structure before and after the adoption of DevOps?*" Later, we primarily focused on the challenges faced by each team, and the practices, decisions, and tools used at the architecture level for adopting DevOps (e.g., "*What architectural characteristics correlate with DevOps?*" The last part of the interviews investigated the architectural decision-making process in the DevOps context (e.g., "*After adopting DevOps in your organization, have you become more independent to make your own (design) decision? If so, how?*").

However, following semi-structured interviews, the participants were allowed to openly discuss any significant DevOps related experiences and insights they had during their respective projects, not limited to the architecture [33]. It is important to mention that we shared the interview guide with the participants before conducting the interviews. This helped them to be prepared for answering the questions and engaging in discussions [33]. The interviews lasted from 40 minutes to one hour and were conducted at the interviewees' workplaces. All six interviews were audio-recorded with the participants' permission and then transcribed, resulting in approximately 40 pages of transcripts.

Besides the interview data, we used more than 120 pages of software development documents provided by the case company to us and publicly available organizational data (e.g., the case company's newsletters). We leveraged those documents to gain a more in-depth insight into important (architectural) decisions made to meet DevOps practices, along with their rationale, as well as to verify experiences and discussions shared by the interviewees. This enabled us to triangulate our findings and increase the validity of our findings [30]. Specifically, for each studied team, we had access to its project plan document, project vision document, architecture document, and internal team discussions forum. All of this data was stored on the case company's wiki and was imported to NVivo[1] software for analysis.

### 2.3 Data Analysis

We analyzed the interview data and the data collected from the documents using open coding and constant comparison techniques from Grounded Theory (GT) [34, 35]. We used NVivo to support qualitative coding and analysis. We first created two top-level nodes in NVivo according to our data sources: (1) interview data and (2) document data. Subsequently, we performed open coding with several iterations in parallel with data collection to thoroughly analyze the data gathered from each data source. This enabled us to identify key points in our data sources and assigning a *code* to each key point. Figure 1 depicts an example of applying open coding on a portion of an interview transcript.

Then, the constant comparison was performed to compare the codes identified in the interviews against each other, as well as to compare them with the codes or the excerpts taken from the documents [35]. We then iteratively grouped these emergent codes to generate *concepts* [34]. Then, the *concepts* created in the previous step were analyzed to develop *categories*, which became the architectural decisions presented in the Findings section. As data analysis progressed, we constructed relationships among the categories.

---

**A portion of an interview transcript**

**Raw data**: "*We're trying to make sure everything [to be] more substitutable, which allows to do mocking if we need it. We're trying to keep everything independent as you can just test that set of function; that succeeded in unit tests*".

**Key point**: "Independent stuff as can be mocked and can be independently tested"

**Code**: Independent units for test

---

**Figure 1**. Constructing codes from the interview transcripts

### 3 Findings

We identified eight high-level architectural decisions of DevOps transformation. It is worth mentioning that software architectures in the context of DevOps not only deal with "*context and requirement*" and "*structure*" aspects of software but also should concern about the "*realization*" aspect [36]. The "*realization*" aspect mostly deals with the operational aspects of software [36] [37]. Moreover, Bass argues [38] that adopting DevOps necessitates organizational, cultural, and technical changes, in all of which software architect plays a key role. This expands the role and responsibility of architects, as they also need to deal with infrastructure architecture, test architecture, team organization, and automation [18] [37] [36]. Hence, the architectural decisions of DevOps transformation in the case company go beyond the application design, and target application architecture, DevOps pipeline architecture, infrastructure architecture, and team organization.

We present the identified decisions using a decision template, including *concern*, *decision*, *implication*, and *technology option* [39]. A *concern* describes the problem that a decision solves. "+" and "-" signs show the positive and negative consequence of a decision on an item respectively (e.g., a decision might influence one or more system quality attributes [3, 40]). "+/-" shows a decision might have both positive and negative consequences. Important consequences are **bold**. Finally, we show the technologies, tools, and tactics that help implement or complement a decision with *technology option* (See Tables 2 to 9 and Tables 11 to 12). It should be noted that when we refer to data from the interviews with TeamA and TeamB, we use **PAX** and **PBX** notations respectively. For instance, **PA1** refers to interviewee 1 in TeamA (See Table 1). The excerpts taken from the documents are marked as **D**.

*Decision 1. External configuration*. We have found that the key architectural decisions in DevOps transformation were about configuration. From the interviews and documents, it is clear that the concept of "external configuration" is at the center of our findings. This can be illustrated by the following quote:

> "*The key architectural thing is [to] making application externally configurable, as you can be aware of the test environment, production environment, etc.*" **PB2**.

External configuration aims at making an application externally configurable, in which each application has embedded configuration (i.e., configurations are bundled within an application) for development and test environments and uses the configuration that lives in the production environment. Here *configuration* refers to storing, tracking, querying, and modifying all artifacts relevant to a project (e.g., application) and the relationships between them in a fully automated fashion [41]. The external configuration also implies **multiple-level configuration** as each environment has a separate configuration. All this makes deploying applications to different environments trivially easy as there is no complicated setup procedure. By this approach, different instances of one artifact in different environments are considered as the same artifact; once one artifact is deployed to the test environment, the same artifact gets deployed into the production environment. The solution architect from TeamA said:

> "*We externalize its [application's] configuration as you can provide different [instances] in different environments but as the same artifact*" **PA2**.

Apart from the positive impact on **deployability**, this decision leads to improving **configurability**. This is because the embedded configurations inside applications can be easily overwritten at target environments, and there is no need to reconfigure the whole infrastructure. Those configurations that might rapidly change are read from Zookeeper[2], but large and static ones are read from HDFS[3] (Hadoop Distributed File System). **PB2** explained the benefits of "external configuration" decision vividly:

> "*We also use Spring Boot. It is always looking for module name first, then for the file directly next to JAR, and then for the embedded JAR. We have multiple levels of configuration in our JAR file. So, inside our JAR, we have the same default, and they always target the local environment. If you*

---

[1] http://www.qsrinternational.com
[2] https://zookeeper.apache.org/
[3] https://hadoop.apache.org/docs/r1.2.1/hdfs_design.html

*accidentally run a JAR file and you might do something crazy like delete data, it lonely targets your local environment".*

Table 2. Decision template for Decision 1

| Concern | Different instances of an application/service should be treated as the same artifact in different environments. |
|---|---|
| Decision | D1. External configuration |
| Implication | **Configurability** (+), Deployability (+), Co-existence (+) |
| Technology | Zookeeper, Hadoop HDFS, Spring Framework |

*Decision 2. Smaller, more loosely coupled units.* Our participants confirm that a DevOps-driven architecture needs to be loosely coupled to ensure team members working on would have a minimum dependency. By this, developers can get their works (e.g., development) done with high isolation [1]. Hence, both teams (A&B) consider "smaller, more loosely coupled units" as a foundation for architecting in DevOps. For example, we have:

> "I guess [we are] trying to be mindful of reducing coupling between different parts of the application as we can separate those things" **PB1**.

> "[In this new architecture,] we want to be able to address each of these stories, without being tightly coupled to another" **D**.

The interviewees reported that this decision is a substantial benefit to **testability** and **deployability**. If decoupled architecture is fulfilled, it appears that team members can independently and easily test units (e.g., better test coverage) and drastically decrease the cycle time of test and deployment processes. **PA1** affirmed that they were successful in implementing this principle as everything for "unit tests" is independent and substitutable, which allows them to do mocking if needed. In both projects, the interviewees explained that the decoupled architecture was achieved by extensive use of dependency injection[4], feature toggle[5], and building units that are backward and forward compatible. A participant from TeamA put it like this:

> "In terms of architecture, [we] build things that they are decoupled through interfaces, using things like dependency injection, using feature flags; these are [the topics related to] the architecture that we use to support deployability" **PA2**.

This architectural decision enabled the teams to make sure everything is nicely separable, reconfigurable, and extensible. According to the participants, all self-contained applications and components were intentionally designed small to be tested in isolation. As an example, we have:

> "Number one thing in DevOps is to have software to be well tested, which you need to separate concerns into separate components that you can test individual piece of functionality [without] influencing other components" **PB1**.

**PB2** discussed the benefit of "smaller, more loosely coupled units" from a different perspective. According to him, breaking down a large component (called Enricher) into five smaller and independent units enabled them to increase test coverage of each to 90%. **PA3**, on the other hand, pointed out that this was also helpful for having a more efficient DevOps pipeline as large components increase DevOps pipeline's time and are hurdles for having quick and direct feedback from test results.

Table 3. Decision template for Decision 2

| Concern | Application's architecture should allow the team to develop and deliver software more quickly. |
|---|---|
| Decision | D2. Smaller, more loosely coupled units |
| Implication | **Testability** (+), **Deployability** (+), Modularity (+), Team collaboration (+), Pipeline time (+), Modifiability (+/-) |
| Technology | Feature toggle, Dependency injection |

*Decision 3. One monolithic deployment unit vs. multiple deployment units.* Whilst TeamA and TeamB had many common architectural decisions (e.g., external configuration) in the context of DevOps, they deal with the deployment process differently. This significantly impacted on their architectural decisions. TeamA had started with microservices architecture style. Still, once the team felt difficulties in the deployment process, they switched to a monolith to minimize the number of deployment units in operations. This significantly helped them to address the deployment challenges encountered earlier with the microservices architecture style. **PA2** and **PA3** explained it in the following words:

> "We actually started with microservices. The reason was that we wanted to scale out some analysis components across machines. The requirements for the application changed, and this led to [move to] this monolith".

> "We deploy it [the application] as one JAR file".

We found the following reasons for building a monolithic-aware deployment: (i) It is much easier to deploy one JAR file instead of deploying multiple deployment units. This mainly because the deployment units are always required to be locked in the end, as some changes are made to all of them at the same time. (ii) Having multiple deployment units can increase overhead in operations time. When **PA1** was asked why their system (i.e., platform) uses one monolithic deployment, he replied:

> "There was a lot of overhead in trying to make sure that this version is backward compatible with this version of that one".

(iii) It can be difficult to manage changes in the development side, where there are multiple, independent deployment units. In this scenario, TeamA found that it is easier to bundle all units together into one monolithic release. **PA1** added:

> "We are doing monolith. I think that's been slowly changing idea; we were originally going to have several suites of [deployment units] that worked together, but I think to cut down on development effort, we are just going to put all in one tool".

Table 4. Decision template for Decision 3.1

| Concern | The number of deployment units should be minimum in production. |
|---|---|
| Decision | D3.1. One monolithic deployment unit in production. |
| Implication | **Deployment** (+), Supportability (+), Operability (+), Testability (-), Scalability (-), Pipeline time (-) |
| Technology | - |

The opposite scenario happened to TeamB as they had recently re-architected its monolithic system to have multiple deployment units. As expressed by **PB2**,

> "We cannot have a thing like TeamA [as] they have one monolithic JAR, and everything is just in there and just deploy to one post. We are very much more microservices; we like to be far away from that [monolithic deployment] as much as possible".

This new architecture, which is called *micro architecture* by TeamB, is a combination of microservices and monolith approaches. Among others (e.g., parts of TeamB's platform should scale and be tested independently of others), we observed that evolvability

---
[4] https://martinfowler.com/articles/injection.html

[5] https://martinfowler.com/articles/feature-toggles.html

(modifiability) was the main business driver for this transition. A member from TeamB commented:

> "*We used to have one big monolithic application, and changing anything required to the redeployment of the whole application, which interrupted all processes. [It is mainly] because we couldn't change anything really without taking down others*" **PB1**.

TeamB used bounded contexts and domain-driven design [42] as a way to split large domains, resulting in smaller, independent, and autonomous deployment units (e.g., microservices and self-contained applications). TeamB's system includes more than forty libraries, microservices, and autonomous applications that are built and deployed automatically and independently. They are very small, and the intention was to keep them small as each of them should be **single bounded** and should do *only* one specific task. This enabled them to minimize external dependencies. **PB2** reported:

> "*We have seven Ingestion apps running; every single app has very specific things to do, like Twitter Ingestor [app] only does Tweets*".

**Table 5**. Decision template for Decision 3.2

| Concern | Parts of the system should be tested, deployed and scaled independently. |
|---|---|
| Decision | D3.2. Multiple deployment units in production. |
| Implication | **Modifiability** (+), Testability (+), Scalability (+), Deployability (+), Modularity (+), Supportability (-) |
| Technology | Domain driven design, bounded context |

*Decision 4. One massive repository for all units/artifacts vs. one repository per unit/artifact*. Another difference we found between TeamA and TeamB was their decision about the repository. Driven by the team's intention to build, test, and deploy the artifacts in isolation, TeamB decided to build and maintain one repository per each artifact. They were frustrated by keeping all artifacts into one repository because with the monolithic repository, changing one thing required running all tests and re-deploy all things. **PB3** referred to this decision as a key architectural decision that was deliberately made to simplify DevOps transformation. This helped them manage the versioning of different libraries, modules, and self-contained applications in the deployment process. Interviewee **PB1** had a similar opinion to **PB3** and said:

> "*The number one thing is that reusable libraries or components should potentially be in own artifact. So, you can build, test, and deploy that artifact in isolation. With the monolith [repository], you change one thing to test that, the monolith is going to run all tests for everything for being (re) deployed*".

We discovered that the concern about the time taken by the DevOps pipeline was another reason for this decision. **PB3** described it as "*the build cycle in our old architecture was problematic as it hit the development and iteration*". In the previous architecture, *only* the build cycle took around 10 minutes. By applying the above changes, the cycle time of the pipeline turned around as currently all the build time, Ansiblizing the deployment process, and release to Nexus[6] take approximately 10 minutes.

**Table 6**. Decision template for Decision 4.1

| Concern | Build, test and deployment of each artifact should be performed in isolation. |
|---|---|
| Decision | D4.1. One repository per unit/artifact |
| Implication | **Pipeline time** (+), Testability (+), Deployability (+), Modifiability (+), Team collaboration (+), Versioning artifacts (+) |
| Technology | - |

Unlike TeamB, TeamA uses one repository for all modules and libraries. This decision was heavily influenced by its monolithic design. The project of TeamA used to have multiple repositories for some of its components and libraries. This caused challenges in the deployment process, such as it made it harder to synchronize artifact' changes, and it presented the integration problem. As explained by **PA3**,

> "*If you make a change to one component, it is still built and passed the tests, but when they would need to be together, it wouldn't work. So, we merged all the components in the same repository*".

**Table 7**. Decision template for Decision 4.2

| Concern | Managing repository should be aligned with the architecture design. |
|---|---|
| Decision | D4.2. One massive repository for all units/artifacts |
| Implication | **Operability** (+), Modifiability (+/-), Supportability (+) |
| Technology | - |

*Decision 5. Application should capture and report status properly*. Monitoring and logging were identified and discussed by all interviewees as a primary area, which needs more attention in the DevOps context. In both projects, they build logs and metrics data into software applications as monitoring tools could leverage these data. Consequently, the applications in the DevOps context need to capture and report data about their operations properly [43]. The system architect in TeamA (**PA3**) believed that this is the most important implication driven by DevOps in terms of architecture. Another interviewee described the relationship between architecture and monitoring as follows:

> "*In order to trust your ecosystem works especially when you are changing things to be continuously deployed; if you don't have strong architecture, then the application' changes are less clear and is more difficult to monitor*" **PB1**.

Both TeamA and TeamB use two monitoring infrastructures, Consul and Ganglia, which are shared among a couple of projects in the case company. These systems are used for different purposes. Consul[7] is mostly used as a high service to check the state of an application. This monitoring system aggregates and orchestrates metrics for cluster state, with which both the teams can identify components' changes among *critical*, *warring*, or *good* states. According to an interviewee, this enables them to prioritize stability issues in clusters, and from that, they are able to implement new fixes and deploy those fixes and capture the relevant information. In contrast, Ganglia[8] is used for aggregating metrics like disc usages, input and output network, interfaces, memory usage, CPU usage. A software engineer in TeamB described the function of Ganglia system as follows:

> "*It [Ganglia] is kind of aggregating the information. For example, this cluster node is constantly 100% CPU usage; this cluster almost is full disc usage. We can identify those stability issues as well and find out them early before becomes a real problem*" **PB3**.

It appears that whilst both the teams extensively improve their logging and monitoring capability using Consul and Ganglia to become DevOps, they are *not* interested in metrics data analytics. Apparently, the log and metrics data are mostly used for diagnostic purposes, and they are not used for an introspective purpose (i.e., understanding how an application/service can be improved based on runtime data). We found that the low number of end users was the reason why TeamA and TeamB are not interested in metric analysis. A participant told us:

> "*We don't do a lot of analysis of the metrics. We do collect a lot of metrics, [for example] we have some centralized logging and centralized monitoring servers and pull all stuff in there, but it is mostly for diagnostic if the application fails and trying to figure out what went wrong*" **PA1**.

---

[6] https://www.sonatype.com/nexus-repository-sonatype
[7] https://www.consul.io/
[8] http://ganglia.sourceforge.net/

In both projects, the abundance of monitoring and logging data produced by the applications presents severe challenges, in which scaling up machines and capacities cannot solve the problem anymore. As a result, most of the participants felt that they have to re-think their logging and monitoring approaches and how to reduce the size of logs. One participant complained about this pain point as follows:

> "*We've got Consul to let us know memory usage in that computer. We've got a server, which executed by multi-tenant solution, and each of them is running their own application that may spin up in parallel; that consumes memory. So, you get the capacity issues. So maybe the solution was to scale out machine initially, but now I think that is an issue again*" **PB1**.

Table 8. Decision template for Decision 5

| Concern | How observability of an application can increase in DevOps context. |
|---|---|
| Decision | D5. Application should capture and report status properly. |
| Implication | **Monitorability** (+), Loggability (+), Supportability (+), Resources utilization (+), Analyzability (+), Cost (-) |
| Technology | Consul, Ganglia |

*Decision 6. Application should be standalone, self-contained.* We found frequent references in the interviews to how the applications can get easily and reliably deployed to different environments. Besides applying "external configuration", both the teams decided that the deployment of an application or service be locked and independent of other applications and services that it depended on. To this end, they have adopted a **self-contained** and **self-hosting** approach [44], in which all dependencies need to be bundled to an application. An interviewee described the impact of this approach as:

> "*Each of them [applications] like Ingestors are considered as a single standalone, self-contained application, and the pipeline is a self-contained app. They don't really talk with any things else; they don't have any external dependencies like taking one of those down doesn't affect the other ones*" **PB2**.

To avoid the interruptions of self-contained applications at runtime, they provide, where possible, the necessary infrastructures (e.g., load balancers) and multiple instances of applications or services behind them. Using Apache Kafka[9] also enables them to segregate service components and isolate them as a service component can be easily swapped in/with new ones at operations:

> "*By using Kafka Buffer, you take a component out and then message its Buffers to either redeployed artifacts or new artifacts that are connected and consume messaging*" **PB3**.

Ansible[10] tool is used as the main automation deployment tool to make sure everything is deployable using infrastructure-as-code. Our interviewees reported two main methods to achieve this. First, it should be ensured that everything, along with its dependencies that are going to be deployed, should be captured in Ansible. Second, Ansible playbooks should be written *well*. From TeamB's architect perspective, writing good Ansible playbooks was the biggest challenge in their DevOps journey.

In TeamB's project, everything is *stateless* as well. In addition, data fields are designed *optional*, which they can either exist or null. This was indeed helpful to the deployment process because there is no need to have the right database with the right data in it to restart the self-contained applications. TeamB extensively uses Zookeeper[11] tool to make code stateless. **PB2** explained:

> "*There is a clean separation between the apps. So, I can go and update Ingestors; I can go and updates DTO and add more data or meta-data to it without affecting anything downstream; like I just continue operating on the old version of schema*".

Table 9. Decision template for Decision 6

| Concern | Application in DevOps context should have the least dependency with others in production (i.e., operational efficiency). |
|---|---|
| Decision | D6. Application should be standalone, self-contained. |
| Implication | **Deployability** (+), Replaceability (+), Recoverability (+), Availability (+), Team effort (-), Cost (-) |
| Technology | Zookeeper, Kafka, Ansible |

*Decision 7. Distinct environments to support different levels of development and deployment processes.* Providing sufficient physical resources (e.g., CPU and memories) to both the teams enabled them to establish three distinct environments, including development, integration (or test), and production environments. In TeamA and TeamB, this was indeed a conscious decision to manage different levels of development, testing, and deployment in DevOps transformation [3].

Table 10. How often the teams deploy to different environments

| | Deployment Frequency to | | |
|---|---|---|---|
| | Development Env. | Integration Env. | Production Env. |
| TeamA | Multiple-time a day | Once per week | Once per Sprint (two-week) |
| TeamB | Multiple-time a day | Multiple-time per | At least once per Sprint |

Table 10 shows the frequency of deployments in these environments is different. They strictly follow three upfront rules in the above-mentioned environments to ease DevOps adoption: (i) in the development environment, ***only*** unit tests need to be performed, and the integration environment should include ***all dependencies*** to properly run integration tests against all snapshot builds. One participant described the test environment like this:

> "*You can bring to your own laptop all the infrastructure requirements of the cluster like Kafka Buffer, Postgre databases, and Zookeeper instances. Then you locally test against that, and then Jenkins automatically tests against the test environment*" **PB1**.

(ii) Code reviews should be performed ***before*** committing to the development branch. TeamA and TeamB use the Gitflow branching model[12], in which developers do their work on feature branches, and before merging changes to the development branch, they should raise a pull request to review the changes in Bitbucket[13] [45]. Then that pull is merged into the development branch. Except for large tasks that might not be merge-able in few days, multiple pull requests occur a day. The interviewee **PA2** reported that these rules "*improve the merge [quality], and [make]*

---

[9] https://kafka.apache.org/
[10] https://www.ansible.com/
[11] https://zookeeper.apache.org/

[12] https://www.atlassian.com/git/tutorials/comparing-workflows/gitflow-workflow
[13] https://bitbucket.org/

*development branch builds always ready for snapshot and deploying to the test environment*".

(iii) Critical bugs should be fixed *only* on the release candidate branch, not on the development branch. This working style, along with automation support, was deemed helpful to faster repair bugs and reduce the risk of deployed changes to production. The interviewee **PB3** said:

> "*If there is an issue, we can quickly fix and redeploy quickly. So, the turnaround is very small because all are automated at the moment*".

Applying the above rules leads to the development branch being potentially in the releasable state anytime, but it is not necessarily stable. An interviewee summarized it as:

> "*It [main branch] is a stable artifact. The development branch, which is all integrated features, ready to be merged into master anytime, [but] that might not be quite stable*" **PB3**.

Although teams deploy to production at the end of a Sprint, actual production deployments are *not* tied to the Sprint and can be frequent.

> "*We are working two weeks Sprint. We always try to our releasable is done in two weeks Sprint but often [we have] more releases, quite often. I think it is releasable in a couple of times a week*" **PB1**.

> "*Releases are not tied to the Sprint tempo - they can be more or less frequent*" **D**.

**Table 11**. Decision template for Decision 7

| Concern | How development and deployment processes can be easier for DevOps context. |
|---|---|
| Decision | D7. Three distinct environments are provided to support different levels of development, testing, and deployment |
| Implication | **Supportability** (+), Testability (+), Deployability (+), Cost (-) |
| Technology | - |

*Decision 8. Teams should be cross-functional, autonomous.* Another key decision made by the studied organization to support DevOps transformation is to set up small, cross-functional teams. Both of the studied teams have **end-to-end ownership** of their products, from design to deployment to support the code in production:

> "*The significant part of the engineering team's time is to maintain the DevOps aspect of the app– I mean the complexity of operations as we have to maintain infrastructure as code and understand that, we have to spend time looking at metrics*" **PB1**.

Furthermore, each team member should perform **testing** and **operations tasks** (e.g., writing Ansible playbooks). In both the teams, testing is considered as a rotating activity, not as a phase, which should be performed by all developers. It should be noted that some software engineers have been frustrated with doing operations tasks, as they believe this can be a source of distraction for doing real software development tasks (e.g., debugging). As an example, **PB2** complained:

> "*Sometimes you [as a developer] spend a lot of time on debugging Ansible and Jenkins and infrastructures things, for example why Kafka Buffer is not coming from this host. It is kind of being dust by DevOps on deployment task*" **PB2**.

Both teams are also **autonomous** in terms of having the freedom to act independently and make their own architectural and technology decisions. For instance, as discussed earlier, they have chosen two different approaches for architecting their respective systems. **Collaboration culture** is well established in both teams through **co-design** and **shared responsibilities**. Regarding the partnership in testing activity, the interviewee **PA1** added:

> "*We are rotating testing role, so one person is responsible for testing everything in the previous Sprint and then press proof button and say: to get deployed to the production*".

Whilst each team has an architect inside it, other team members actively and substantially participated in the architecting process. The architects are mainly responsible for designing and evaluating systems; however, they also code and test. The architect of TeamA put the collaborative design in these words:

> "*Everybody is involved in architecture [design]. We are not going to have only I as a solution architect; everyone in the team can do that. The architecture is done in daily work, and everyone can have his [own] idea*" **PA2**.

**Table 12**. Decision template for Decision 8

| Concern | How teams should be reorganized to effectively adopt DevOps. |
|---|---|
| Decision | D8. Teams should be cross-functional, autonomous. |
| Implication | **Team collaboration** (+), Team effort (-) |
| Technology | Confluence, HipChat, JIRA, Bitbucket |

## 4 Lessons Learned

In the previous section, we have provided the key insights about architectural decisions and their potential implications for the DevOps paradigm in our case study. In the following, we discuss some of the key lessons learned in the context of the related work and the implications of our results, which are expected to be helpful for other organizations and practitioners trying to adopt DevOps.

**Modularity is the key:** Whilst some organizations [21, 22] have adopted microservice-based architecture as a driver to succeed in DevOps, our results have revealed that achieving DevOps-driven architectures requires loosely coupled architectures and prioritizing deployability, testability, supportability, and modifiability over other quality attributes (i.e., confirming and extending [18, 20]). This is because improving/addressing these quality attributes was the target of the majority of the identified decisions. For example, deployability was positively influenced by seven *Decisions* 1, 2, 3.1, 3.2, 4.1, 6, 7. Our findings are in line with the *2017 State of DevOps Report* [1], which revealed that apart from architecting a system with microservices style, service-oriented architecture or any other architecture styles, loosely coupled architectures and teams are the most significant contributors to DevOps success (i.e., releasing software changes at high velocity). Etsy is a notable example of this scenario, as it has successfully implemented DevOps using a monolithic system [46]. Whilst both TeamA and TeamB realized that the fundamental limitation to rapid and safe delivery resides in their systems' architectures, we *did* find significant differences in their architectural approaches: monolith vs. micro architecture. Here monolith means a single build system, in which all the functionality is managed and deployed in one deployable unit [47, 48]. This does not necessarily mean that a monolith is a highly coupled architecture. TeamA uses one monolithic deployment unit (i.e., *Decision* 3.1), whilst TeamB deploys several deployment units in operations (i.e., *Decision* 3.2).

We observed that TeamB is more likely to have loosely coupled architecture than that of TeamA. It is mainly because changing the architecture of TeamA's platform is not straightforward, as TeamA's architect needed to get involved frequently whenever the architectural changes required. Conversely, TeamB members could apply small-scale changes to their architecture without the need for their architect. However, in both systems applying large-scale changes required the involvement of their architects. Another sign that shows the architecture of TeamB's platform is more modular is that TeamB had higher deployment frequency to lower environments (e.g., test environment) compared to TeamA (See Table 10). Other studies [18, 49] reveal that the architectures in the DevOps context should support evolutionary changes. This implies that architectural decisions should be delayed until they are necessary (i.e., delaying decisions to the last possible moment) [49]. However, it does not mean that there is no longer a need to have a long-term vision of architecture. Put another way, core architectural decisions need to be made at the early stage of development in the DevOps context. By ignoring this necessity, TeamA experienced a pain point in the architecting process as they ended up major refactoring of the whole stack of their system several times.

**Better to add operations specialists to the teams**: Our study shows that establishing cross-functional, autonomous teams was one of the key decisions made in the case company to implement DevOps (i.e., _Decision_ 8). We observed that operations expert is not embodied as a distinct role on both TeamA and TeamB, or there is no dedicated operations team in the case company. The operations skills are regarded as a skillset that blends with other skills such as software development, and the operations tasks need to be performed by all members. We observed that performing the operations tasks that require deep expertise in operations (e.g., writing Ansible playbook) is burdensome for the teams. This caused a significant part of the team members' time is spent on the operations tasks rather than the development tasks (i.e., the negative consequence of _Decisions_ 6 and 8 on team effort). We also found that that this can give a good excuse for the teams to do *not* perform the operations tasks optimally. This can be exemplified by the Ansible playbooks written by the team's members as part of their new responsibilities in DevOps. Whilst the teams' members found that the quality of the written Ansible playbooks is not good enough, we found that there is not any demand to improve them. In our view, this could stem from the fact that operations responsibilities are ambiguous for the teams' members, and some of the operations tasks are not clear who should do. Furthermore, in the DevOps transformation, the development side is more emphasized than the operations side [18]. This could be mainly justified by the fact that most of the business values come from the development side (e.g., adding more features). We emphasize that operations specialists need to be embedded in each team for the complex operations tasks [50, 51]. This would be beneficial for development people as they can concentrate more on real software engineering tasks. This is in line with the findings of [15], which showed that the majority of the surveyed organizations (76 out of 93) prefer having distinct operations team or operations specialists for specific operations tasks.

**The Road Ahead (Remaining Challenges)**: One of the most emphasized DevOps practices is Continuous Delivery or Deployment (CD) [38]. Despite having an automated deployment to production and designing with deployability in mind, both teams *are not able* to or *do not want* to practice CD (i.e., release happens once in two weeks). Whilst truly implementing CD might be influenced by many socio-technical factors (e.g., in the case company: the circumstance of the projects), we observed a strong influence of test automation on the deployment speed. Both teams have testing-related problems, including *lack of having good test coverage*, *lack of comprehensive end-to-end integration test automation*, and *conducting performance, security, and acceptance tests out of DevOps pipelines*. Given the size of the teams and end users, and the overhead (i.e., time and cost) of developing and maintaining automated performance and acceptance tests, both teams prefer to do these sorts of tests manually. For instance, as the overhead of maintaining Selenium-based User Interface (UI) testing increased (e.g., because UI changes a lot), TeamA found that it is better to turn off UI testing in the deployment pipeline. That is why currently, they do it manually on the release branch.

All these issues have reduced the studied teams' confidence in deploying multiple times a day. Many architectural decisions (i.e., _Decisions_ 2, 3.1 3.2, 4.1, 7) reported in Section 3 were made to improve the testability of an application in the DevOps context; however, we argue that organizations should also ensure to have good test coverage, write tests that less consume cycle time of a DevOps pipeline, and automate tests (e.g., performance) that occur during the last stages of DevOps pipelines for successfully implementing CD [2] [38]. This provides confidence to deploy to production continuously and automatically.

## 5 Threats to Validity

Our research method and findings in this case study may have a few limitations, which we discuss below from the qualitative research perspective [52, 53].

**Transferability**: The main limitation of this study is that the findings (e.g., architectural decisions and design challenges) are entirely based on one organization in a particular domain with a small sample size. While we have tried to minimize the potential validity impact of this limitation by studying two independent teams working on two different projects in one organization, our findings may not be (statistically) generalizable to other organizations or domains. However, it should be noted that the focus of this study was to provide a deep understanding of architectural decisions, tactics, and problems within the case company in DevOps transformation. Moreover, this study aimed at discussing important lessons learned through rigorous data collection and analysis [30], so that other organizations and practitioners can benefit from that. Finally, it is worth mentioning that the studies involving a single organization are regarded as valuable contributions in the software engineering community, as they contribute to scientific development [54] [55]. For example, the studies [56] [57] [58] are based on single case studies collected data from a small sample size: 10 interviews [56], 6 interviews [57], and 5 interviews (complemented with observation) [58].

**Credibility**: Using two different data sources (i.e., interview and documentation) and investigating two different teams ensured that the obtained findings to a large extent are plausible. The selection of the participants can be another threat to the credibility of our findings. To recruit motivated participants, we ensured that personal details, opinions, and thoughts would be regarded as strictly confidential and would not be divulged to the researchers and other team members by making a confidential disclosure agreement. After discussing the objectives of this study with the CTO and two key informants at the case company, the CTO introduced us to a few suitable persons from each team for the interviews who were invited for participation. This gives us confidence that the team members who chose to participate were likely more willing and had the right types of competencies to provide an unbiased opinion.

Another possible limitation comes from the interview questions. Our interview questions were primarily designed based on a comprehensive systematic review [2] and the existing empirical studies on software architecture and DevOps [1] [18] [20]. In addition, we tried to fine-tune the questions and asked appropriate follow-up questions according to the participants' responses and their projects. Another threat to the credibility of our study is whether the reported architectural decisions were exclusively driven by DevOps transformation, not because of other confounding factors. We accept that there might be decisions that do not strictly concern DevOps or could be applied to other software development methodologies (e.g., *Decision 1* is like one of best practices in twelve-factor methodology [59] or migrating applications to the cloud [60]). To mitigate this threat, we took two actions: in the meeting with CTO and two informants, we described the objective of our study. In addition to this, during the interviews, we encouraged the interviewees to mainly discuss the changes in the architecting process in the context of DevOps.

**Confirmability**: Data analysis was conducted by one researcher (i.e., the first author). Whilst this helped to obtain consistency in the results [61], it can be a potential threat to the validity of the findings. This threat was mitigated to some extent by organizing internal discussions to review and verify the findings and solicit feedback. In addition to this, we minimized the subjective bias by asking the CTO of the case company to review and provide feedback on the early version of this paper. Furthermore, other research [53] highlights that data triangulation strategy, which previously described for improving credibility, can be used to establish confirmability as it can reduce the subjectivity of the researcher's understanding and judgment.

## 6 Related Work

This section presents existing research that has investigated the role of software architecture in the context of DevOps. Over the last eight years, Puppet[14] has annually released non-peer reviewed reports to study the current state of DevOps in practice [1, 62]. In 2017 [1], the role of software architecture in DevOps was deeply examined to investigate how application architecture, and the structure of the teams that work on, impact the delivery capability of an organization. The main finding of this report reads, "loosely coupled architectures and teams are the strongest predictors of continuous delivery". Surprisingly, it was found that many so-called service-oriented architectures (e.g., microservices) in practice may prevent testing and deploying services independently from each other. Subsequently, it can negatively influence teams to develop and deliver software.

Shahin et al. [18] have conducted a mixed-methods study to explore how SA is being impacted by or is impacting CD. They present a conceptual framework to support (re-) architecting for CD. The work by Mårtensson et al. [63] identified 12 enabling factors (e.g., "work breakdown" and "test before commit") impacting the capability of the developers when practicing CI. The study [63] argues that some of these factors (e.g., work breakdown) can be limited by the architecture of a system. Shahin et al. [18], Chen [20], Chen [23] argue that a set of quality attributes such as deployability, security, modifiability, and monitorability, require more attention when designing architectures in the context of CD. Shahin et al. [18] also found that a lack of explicit control on reusability poses challenges to practicing CD. Cukier [64] applied eight design patterns (e.g., "queue-based solution to process asynchronous jobs") to design the architecture of a system in a DevOps context. It was observed that applying those patterns led to a more loosely coupled architecture. The studies [18, 65] found that monoliths are the main source of pain to practice CD in the industry. In a retrospective study on three projects adopting continuous integration and delivery, Bellomo et al. [19] revealed that the architectural decisions made in those projects played a significant role in achieving the desired state of deployment (i.e., deployability). Di Nitto et al. [66] outlined architecturally significant stakeholders (e.g., infrastructure provider) and their concerns (e.g., monitoring) in DevOps scenarios. Then a framework called SQUID was built, which aims at supporting the documentation of DevOps-driven software architectures and their quality properties. In [67], the Business Process Architecture modeling technique was utilized to visualize different aspects of DevOps process and to help organizations to customize their DevOps processes based on their needs and constraints. Based on an experience report, Balalaie et al. [21] presented the architectural patterns (e.g., change code dependency to service call) and technology decisions (e.g., using containerization to support continuous delivery) employed by a case company to re-architect a monolithic architecture into microservices in the context of DevOps. They also reported that a microservices-based architecture might pose some challenges, such as bringing higher complexity. In [22], Callanan and Spillane discuss that developing a standard release path and implementing independently releasable microservices through building backward compatibility with each release were the main tactics leveraged by their respective company to smooth DevOps transformation. These tactics also significantly reduced delays in the deployment pipeline.

## 7 Conclusion

We have described the results of a case study aimed at learning about the architectural decisions made and the challenges faced by two teams when architecting their systems as part of their DevOps journey. Our findings suggest that DevOps success is best associated with modular architectures and needs to prioritize deployability, testability, supportability, and modifiability over other quality attributes. We believe that the successful architectural decisions made by the teams to support DevOps will be valuable for other organizations. Our observations from the current setting of the studied teams show that the teams can improve their performance in DevOps by (1) adding operations specialists in the teams to perform the operations tasks that require advanced expertise and (2) investment in testing, in particular automating tests occur during the last stages of DevOps pipelines, to release software changes faster. In the future, we plan to validate and generalize our findings by conducting case studies in other organizations. This would enable us to develop a decision guidance template for DevOps-driven architectures, which can ensure the architecture of a software system is natural to DevOps.

### Acknowledgment

When the main part of this work was carried out, the first author was affiliated with University of Adelaide and supported by Australian Government Research Training Program Scholarship and Data61 (a business unit of CSIRO). The authors would like to thank the participants and the CTO of the case company.

### References

[1] N. Forsgren, G. Kim, N. Kersten, J. Humble, and A. Brown, *2017 State of DevOps report*, 2017.
[2] M. Shahin, M. A. Babar, and L. Zhu, "Continuous Integration, Delivery and Deployment: A Systematic Review on Approaches, Tools, Challenges and Practices," *IEEE Access*, vol. 5, pp. 3909-3943, 2017.
[3] L. Bass, I. Weber, and L. Zhu, *DevOps: A software architect's perspective*: Addison-Wesley Professional, 2015.

---

[14] https://puppet.com/

[4] L. Leite, C. Rocha, F. Kon, D. Milojicic, and P. Meirelles, "A Survey of DevOps Concepts and Challenges," *ACM Computing Surveys (CSUR)*, vol. 52, no. 6, pp. Article 127, 2019.
[5] A. A. U. Rahman, and L. Williams, "Software security in devops: synthesizing practitioners' perceptions and practices," in IEEE/ACM International Workshop on Continuous Software Evolution and Delivery (CSED), 2016, pp. 70-76.
[6] M. G. Jaatun, D. S. Cruzes, and J. Luna, "Devops for better software security in the cloud invited paper," in 12th International Conference on Availability, Reliability and Security, 2017, pp. 1-6.
[7] S. Mäkinen, M. Leppänen, T. Kilamo, A.-L. Mattila, E. Laukkanen, M. Pagels, and T. Männistö, "Improving the delivery cycle: A multiple-case study of the toolchains in Finnish software intensive enterprises," *Information and Software Technology*, vol. 80, pp. 175-194, 2016.
[8] H. Kang, M. Le, and S. Tao, "Container and microservice driven design for cloud infrastructure devops," in IEEE International Conference on Cloud Engineering (IC2E), 2016, pp. 202-211.
[9] J. Wettinger, U. Breitenbücher, O. Kopp, and F. Leymann, "Streamlining DevOps automation for Cloud applications using TOSCA as standardized metamodel," *Future Generation Computer Systems*, vol. 56, pp. 317-332, 2016.
[10] A. van Hoorn, P. Jamshidi, P. Leitner, and I. Weber, "Report from GI-Dagstuhl Seminar 16394: Software Performance Engineering in the DevOps World," *arXiv preprint arXiv:1709.08951*, 2017.
[11] L. E. Lwakatare, T. Kilamo, T. Karvonen, T. Sauvola, V. Heikkilä, J. Itkonen, P. Kuvaja, T. Mikkonen, M. Oivo, and C. Lassenius, "DevOps in Practice: A Multiple Case study of Five Companies," *Information and Software Technology*, 2019.
[12] W. P. Luz, G. Pinto, and R. Bonifácio, "Adopting DevOps in the real world: A theory, a model, and a case study," *Journal of Systems and Software*, vol. 157, pp. 110384, 2019.
[13] F. Erich, C. Amrit, and M. Daneva, "A qualitative study of DevOps usage in practice," *Journal of Software: Evolution and Process*, vol. 29, no. 6, pp. e1885, 2017.
[14] L. E. Lwakatare, T. Karvonen, T. Sauvola, P. Kuvaja, H. H. Olsson, J. Bosch, and M. Oivo, "Towards devops in the embedded systems domain: Why is it so hard?," in Hawaii International Conference on System Sciences (HICSS), 2016, pp. 5437-5446.
[15] M. Shahin, M. Zahedi, M. A. Babar, and L. Zhu, "Adopting continuous delivery and deployment: Impacts on team structures, collaboration and responsibilities," in 21st International Conference on Evaluation and Assessment in Software Engineering, 2017, pp. 384-393.
[16] K. Nybom, J. Smeds, and I. Porres, "On the impact of mixing responsibilities between devs and ops," in International Conference on Agile Software Development, 2016, pp. 131-143.
[17] M. Skelton, and M. Pais, *Team Topologies: Organizing Business and Technology Teams for Fast Flow*: IT Revolution, 2019.
[18] M. Shahin, M. Zahedi, M. A. Babar, and L. Zhu, "An empirical study of architecting for continuous delivery and deployment," *Empirical Software Engineering*, vol. 24, no. 3, pp. 1061-1108, 2019/06/01, 2019.
[19] S. Bellomo, N. Ernst, R. Nord, and R. Kazman, "Toward design decisions to enable deployability: Empirical study of three projects reaching for the continuous delivery holy grail," in Annual IEEE/IFIP International Conference on Dependable Systems and Networks, 2014, pp. 702-707.
[20] L. Chen, "Towards architecting for continuous delivery," in Working IEEE/IFIP Conference on Software Architecture, 2015, pp. 131-134.
[21] A. Balalaie, A. Heydarnoori, and P. Jamshidi, "Microservices architecture enables devops: Migration to a cloud-native architecture," *Ieee Software*, vol. 33, no. 3, pp. 42-52, 2016.
[22] M. Callanan, and A. Spillane, "DevOps: making it easy to do the right thing," *Ieee Software*, vol. 33, no. 3, pp. 53-59, 2016.
[23] L. Chen, "Microservices: architecting for continuous delivery and DevOps," in International Conference on Software Architecture (ICSA), 2018, pp. 39-397.
[24] U. Zdun, E. Wittern, and P. Leitner, "Emerging Trends, Challenges, and Experiences in DevOps and Microservice APIs," *IEEE Software*, vol. 37, no. 1, pp. 87-91, 2019.
[25] XebiaLabs, *Exploring Microservices: 14 Questions Answered By Experts*.
[26] M. Schmidt. "DevOps and Continuous Delivery: Not the Same," https://bit.ly/2vEme4H.
[27] T. Laukkarinen, K. Kuusinen, and T. Mikkonen, "DevOps in regulated software development: case medical devices," in 39th International Conference on Software Engineering: New Ideas and Emerging Results Track, 2017, pp. 15-18.
[28] M. Shahin, M. A. Babar, M. Zahedi, and L. Zhu, "Beyond continuous delivery: an empirical investigation of continuous deployment challenges," in 11th ACM/IEEE International Symposium on Empirical Software Engineering and Measurement, Markham, Ontario, Canada, 2017, pp. 111–120.
[29] L. Zhu, L. Bass, and G. Champlin-Scharff, "DevOps and Its Practices," *IEEE Software*, vol. 33, no. 3, pp. 32-34, 2016.
[30] R. K. Yin, *Case study research and applications: Design and methods*: Sage publications, 2017.
[31] P. Runeson, and M. Höst, "Guidelines for conducting and reporting case study research in software engineering," *Empirical software engineering*, vol. 14, no. 2, pp. 131, 2009.
[32] L. Prechelt, H. Schmeisky, and F. Zieris, "Quality experience: a grounded theory of successful agile projects without dedicated testers," in 2016 IEEE/ACM 38th International Conference on Software Engineering (ICSE), 2016, pp. 1017-1027.
[33] S. E. Hove, and B. Anda, "Experiences from conducting semi-structured interviews in empirical software engineering research," in 11th IEEE International Software Metrics Symposium (METRICS'05), 2005, pp. 10 pp.-23.
[34] B. G. Glaser, A. L. Strauss, and E. Strutzel, "The discovery of grounded theory; strategies for qualitative research," *Nursing research*, vol. 17, no. 4, pp. 364, 1968.
[35] R. Hoda, "Self-organizing agile teams: A grounded theory," Victoria University of Wellington, 2011.
[36] G. Hohpe, I. Ozkaya, U. Zdun, and O. Zimmermann, "The Software Architect's Role in the Digital Age," *IEEE Software*, vol. 33, no. 6, pp. 30-39, 2016.
[37] E. Woods, "Operational: The Forgotten Architectural View," *IEEE Software*, vol. 33, no. 3, pp. 20-23, 2016.
[38] L. Bass, "The software architect and DevOps," *IEEE Software*, vol. 35, no. 1, pp. 8-10, 2017.
[39] S. Haselböck, R. Weinreich, and G. Buchgeher, "Decision guidance models for microservices: service discovery and fault tolerance," in Fifth European Conference on the Engineering of Computer-Based Systems, 2017, pp. 4.
[40] I. O. f. Standardization, "ISO/IEC 25010:2011 Systems and software engineering–systems and software quality requirements and evaluation (square)–system and software quality models," *International Organization for Standardization*, 2011, p. 2910.
[41] J. Humble, and D. Farley, *Continuous Delivery: Reliable Software Releases through Build, Test, and Deployment Automation (Adobe Reader)*: Pearson Education, 2010.
[42] E. Evans, *Domain-driven design: tackling complexity in the heart of software*: Addison-Wesley Professional, 2004.
[43] J. Bosch, "Speed, data, and ecosystems: the future of software engineering," *IEEE Software*, vol. 33, no. 1, pp. 82-88, 2015.
[44] T. Cerny, M. J. Donahoo, and J. Pechanec, "Disambiguation and comparison of soa, microservices and self-contained systems," in International Conference on Research in Adaptive and Convergent Systems, 2017, pp. 228-235.
[45] G. Gousios, M.-A. Storey, and A. Bacchelli, "Work practices and challenges in pull-based development: the contributor's perspective," in 2016 IEEE/ACM 38th International Conference on Software Engineering (ICSE), 2016, pp. 285-296.
[46] D. Schauenberg, "Development, deployment and collaboration at Etsy."
[47] R. Stranghöner. "Self-Contained Systems. Assembling Software from Independent Systems," https://scs-architecture.org/.
[48] N. Dragoni, S. Giallorenzo, A. L. Lafuente, M. Mazzara, F. Montesi, R. Mustafin, and L. Safina, "Microservices: yesterday, today, and tomorrow," *Present and ulterior software engineering*, pp. 195-216: Springer, 2017.
[49] M. Erder, and P. Pureur, *Continuous architecture: sustainable architecture in an agile and cloud-centric world*: Morgan Kaufmann, 2015.
[50] G. Bergman. "Serving 86 million users – DevOps the Netflix way," https://bit.ly/3cLdFFZ.
[51] G. Haff. "DevOps success: A new team model emerges," https://red.ht/2VTppjx.
[52] E. G. Guba, "Criteria for assessing the trustworthiness of naturalistic inquiries," *Ectj*, vol. 29, no. 2, pp. 75, 1981.
[53] K.-J. Stol, P. Avgeriou, M. A. Babar, Y. Lucas, and B. Fitzgerald, "Key factors for adopting inner source," *ACM Transactions on Software Engineering and Methodology*, vol. 23, no. 2, pp. Article 18, 2014.
[54] B. Flyvbjerg, "Five misunderstandings about case-study research," *Qualitative inquiry*, vol. 12, no. 2, pp. 219-245, 2006.
[55] E. Kalliamvakou, C. Bird, T. Zimmermann, A. Begel, R. DeLine, and D. M. German, "What Makes a Great Manager of Software Engineers?," *IEEE Transactions on Software Engineering*, vol. 45, no. 1, pp. 87-106, 2019.
[56] P. Rodríguez, E. Mendes, and B. Turhan, "Key Stakeholders' Value Propositions for Feature Selection in Software-intensive Products: An Industrial Case Study," *IEEE Transactions on Software Engineering*, pp. 1-1, 2018.
[57] M. Senapathi, J. Buchan, and H. Osman, "DevOps Capabilities, Practices, and Challenges: Insights from a Case Study," in Conference on Evaluation and Assessment in Software Engineering, New Zealand, 2018, pp. 57–67.
[58] P. Lous, P. Tell, C. B. Michelsen, Y. Dittrich, and A. Ebdrup, "From Scrum to Agile: a journey to tackle the challenges of distributed development in an Agile team," in 2018 International Conference on Software and System Process, Gothenburg, Sweden, 2018, pp. 11–20.
[59] A. Wiggins. "Twelve-Factor App methodology," https://12factor.net/.
[60] J. Cito, P. Leitner, T. Fritz, and H. C. Gall, "The making of cloud applications: an empirical study on software development for the cloud," in 10th Joint Meeting on Foundations of Software Engineering, Bergamo, Italy, 2015, pp. 393–403.
[61] K. F. Tómasdóttir, M. Aniche, and A. v. Deursen, "Why and how JavaScript developers use linters," in 32nd IEEE/ACM International Conference on Automated Software Engineering, 2017, pp. 578-589.
[62] A. Mann, M. Stahnke, A. Brown, and N. Kersten, *2019 State of DevOps Report*, 2019.
[63] T. Mårtensson, D. Ståhl, and J. Bosch, "Continuous integration impediments in large-scale industry projects," in 2017 IEEE International Conference on Software Architecture (ICSA), 2017, pp. 169-178.
[64] D. Cukier, "DevOps patterns to scale web applications using cloud services," in Proceedings of the 2013 companion publication for conference on Systems, programming, & applications: software for humanity, 2013, pp. 143-152.
[65] G. Schermann, J. Cito, P. Leitner, U. Zdun, and H. Gall, *An empirical study on principles and practices of continuous delivery and deployment*, 2167-9843, PeerJ Preprints, 2016.
[66] E. D. Nitto, P. Jamshidi, M. Guerriero, I. Spais, and D. A. Tamburri, "A software architecture framework for quality-aware DevOps," in 2nd International Workshop on Quality-Aware DevOps, Saarbrücken, Germany, 2016, pp. 12–17.
[67] Z. Babar, A. Lapouchnian, and E. Yu, "Modeling DevOps Deployment Choices Using Process Architecture Design Dimensions," in The Practice of Enterprise Modeling, Cham, 2015, pp. 322-337.